\documentclass[aip,twocolumn,reprint,superscriptaddress]{revtex4-1}
\usepackage{graphicx}
 \usepackage{amsmath}
 \usepackage{amssymb}
 \usepackage{color}
 
\renewcommand{\eqref}[1]{Eq.\,(\ref{#1})}
\newcommand{\figref}[1]{Fig.\,\ref{#1}}
\newcommand{\secref}[1]{Sec.\,\ref{#1}}

\begin{document}

\title{On the band-width of stable nonlinear stripe patterns in finite size systems}

\author{Mirko Ruppert}
\affiliation{Theoretische Physik, Universit\"at Bayreuth, 95440 Bayreuth, Germany}

\author{Walter Zimmermann}
\email[]{walter.zimmermann@uni-bayreuth.de}
\affiliation{Theoretische Physik, Universit\"at Bayreuth, 95440 Bayreuth, Germany}

\date{\today}

\begin{abstract}
Nonlinear stripe patterns occur in many different systems, from the small scales of biological cells to geological scales as cloud patterns.
They all share the universal property of being stable at different
wavenumbers $q$, i.e., they are multistable. 
The stable wavenumber range of the stripe patterns, which is limited by the Eckhaus- and zigzag instabilities even in finite systems for several boundary conditions, increases with decreasing system size.
This enlargement comes about because suppressing degrees of freedom from the two instabilities goes along with the system reduction, and the enlargement  depends on the boundary conditions, as we show analytically and numerically with the generic Swift-Hohenberg (SH) model and the universal Newell-Whitehead-Segel equation. 
 We also describe how, in very small system sizes, any periodic pattern that emerges from the basic state is simultaneously stable in certain parameter ranges, which is especially important for Turing pattern in cells.   In addition, we explain why below a certain system width stripe pattern behave quasi-one-dimensional in two-dimensional systems.
Furthermore, we show with numerical simulations of the SH
model in medium-sized rectangular domains how unstable stripe patterns evolve via the zigzag instability  differently into stable patterns for different combinations of boundary conditions.  
\end{abstract}

\pacs{}
\maketitle


\begin{quotation}
Nonlinear stripe patterns are ubiquitous in nature, and their driving mechanisms 
are as diverse as the systems themselves in which they occur
\cite{CrossHo,Ball:98,Busse:78.1,Kramer:96,Aranson:02.1,Kapral:1995,BoPeAh:2000.1,Mikhailov:2006.1,Pismen:2006,Kondo_Miura:2010.1,Lappa:2010,Sasai:2013.1,Meron:2015,Meron:2018.1,BaerM:2020.1}.
Stripe patterns have the universal property of being stable at different values 
of the wavenumber, and these stable wavenumber regions are the so-called Busse balloons after their pioneer \cite{Busse:1967.2,Busse:78.1,CrossHo,Newell:1993.1,Lappa:2010}. 
To the  instabilities bounding the stable wavenumber range of stripe patterns count the generic Eckhaus instability 
 \cite{ }, a long-wavelength longitudinal (compressional) instability, and
the zigzag instability, a long-wavelength transverse instability \cite{Newell:1969.1,CrossHo}.
Stable wavenumber ranges are restricted even in large systems by pattern suppressing
 boundary conditions at the domain sides 
\cite{CDHS,Kramer:1984.1} or are even selected by spatial inhomogeneities, e.g. via so-called ramps 
\cite{Kramer:82.1,Ahlers:83.1,Cross:1984.1}. In contrast, in short systems the stable wavenumber range can be enlarged, as in Ref. \cite{Zimmermann:85.1} for quasi-one-dimensional systems predicted and experimentally confirmed in Ref. \cite{Ahlers:86.1,Dominguez-Lerma:86.2}.
Such a range extension depends on the boundary conditions and the second spatial dimension as we explain  in this work analytically and numerically by investigating  
the generic Swift-Hohenberg model and the universal Newell-Whitehead-Segel equation
in  rectangular domains. Finite size effects on patterns are also highly relevant for Turing 
patterns in small systems, as for instance in cells 
\cite{deBoer:1999.1,DekkerC:2015.1,Sourjik:2017.1,Bergmann:2018.1}.
\end{quotation}

\section{Introduction}\label{sec: intro} 

Patterns occur spontaneously in a plethora of living or inanimate driven systems, such as in the atmosphere or in convection cells, in biological cells, in chemical reactions, or as vegetation patterns, to name just a few examples
 \cite{CrossHo,Ball:98,Busse:78.1,Kramer:96,Aranson:02.1,Kapral:1995,BoPeAh:2000.1,Mikhailov:2006.1,Pismen:2006,Kondo_Miura:2010.1,Lappa:2010,Sasai:2013.1,Meron:2015,Meron:2018.1,BaerM:2020.1}.
   Already the
  esthetic appeal of  patterns is immediately apparent to all observers \cite{Ball:98}.
 Patterns  fulfill also important functions in nature. For example, 
  self-organized patterns in biology guide size sensing \cite{LanderAD:2011.1}, positioning of protein clusters in the cell center in advance of cell division
  \cite{Sourjik:2017.1} or in  self-driven morphogenesis \cite{Sasai:2013.1}.
   Patterns enhance transport in fluid systems \cite{CrossHo,Lappa:2010} 
  or they are the basis of successful survival strategies for vegetation in water-limited systems \cite{Meron:2015,Meron:2018.1}.  
 
 Patterns are multistable, i.e. they are stable for different wavenumbers within a stability band \cite{CrossHo,Newell:1969.1,Busse:78.1}, sometimes also beyond a seondary instability \cite{ThomsenF:2021.1}. In quasi-one-dimensional systems, the stability band is bounded by the Eckhaus instability
  \cite{Eckhaus:65,Newell:1969.1,CrossHo,Boucif:84.1,Zimmermann:85.1,Ahlers:86.1,Dominguez-Lerma:86.2,Riecke:86.1,Lowe:85.2,Zimmermann:85.3,Dominguez-Lerma:86.1,Zimmermann:88.1,Tuckerman:1990.1}, 
 which also has its two-dimensional generalization in anisotropic systems \cite{Kramer:1986.1, Zimmermann:88.3}. In two-dimensional isotropic systems, the stability band of stripe patterns is in addition bounded by the zigzag instability \cite{Newell:1969.1,CrossHo,Cross:2009}. 
  
  In nature, patterns are always exposed to boundaries, be it the walls of a convection cell, the finite size of a Petri dish, or the cytosol bounding membrane for intracellular processes. Along these boundaries, the fields describing the patterns must satisfy certain boundary conditions. Some boundary conditions  suppress the pattern near the boundary. In longitudinal  direction  they act even in long sytems far into the volume  and significantly restrict the stable wavenumber band \cite{CDHS}. 
In contrast, periodic and no-flux boundary conditions, for example, impose no restriction on the stable wavenumber band, except that the wavenumbers can take only discrete values. In rectangular systems with amplitude-suppressing boundary conditions, stripe patterns prefer an orientation perpendicular to these edges \cite{Greenside:84.1,CrossHo,Ruppert:2020.1}.

Patterns are also restricted to finite ranges when the control parameter generating a pattern falls to subcritical values outside a subrange. Examples include photosensitive chemical reactions. There, pattern formation can be suppressed by illuminating the reaction cell outside a finite range \cite{Epstein:1999.1} or even controlled by spatially modulated illumination
 \cite{Epstein:2001.1,PeterR:2005.1,Hammele:2006.2}. Another example is protein patterns occurring in finite reactive subdomains of substrates \cite{Schwille:2012.1}. In such examples, orientations are often perpendicular to non-resonant control parameter drops \cite{Rapp:2016.1,Ruppert:2020.1}. In the case of steep control parameter decays, stationary strips may orient parallel to the boundaries due to resonance effects \cite{Rapp:2016.1}.
In quasi-one-dimensional systems, spatial variations, called ramps, break the translational symmetry and drastically reduce the width of the stable wavenumber band \cite{Kramer:82.1,Ahlers:83.1,Cross:1984.1,Kramer:1985.1,Riecke:86.1}.

No-flux boundary conditions play a central role for reaction-diffusion patterns in cells and elsewhere.  No-flux boundary conditions break translational symmetry and fix the phase of a periodic pattern at the boundary, but 
 they share the property of leaving the threshold untouched and, in medium sized systems, also the stable wavenumber band. Both boundary conditions are very well suited to study the direct influence of the system size on the stability range of striped patterns. From investigations on quasi one-dimensional systems it is known that 
 a decrease of the system size leads to an increase of the Eckhaus stable wavenumber band \cite{Zimmermann:85.1}, which is also confirmed experimentally \cite{Ahlers:86.1,Dominguez-Lerma:86.2}. This is an opposite trend 
 as obtained for medium sized and long systems with amplitude suppressing boundary conditions in longitudinal direction and for ramps.

Therefore, an interesting question arises: how will the zigzag instability boundary of the stable wavenumber band be shifted by reducing the size of the domain containing the patterns?
We address this question by studying the stability of stripe solutions of the Swift-Hohenberg model in rectangular domains as well as and simultaneously with the universal Newell-Whitehead-Segel equation.  Since in rectangular domains stripe  patterns are oriented perpendicular to edges with amplitude suppressing conditions, we choose either  periodic or  no-flux boundary conditions along the short sides of the rectangle and combine them with these two boundary conditions also along the long sides of the rectangle. This allows us to analytically determine the Eckhaus stability boundary and the zigzag stability boundary in section \ref{sec:Lin_stab_stripes}.
Combining these boundary combinations gives very good analytical estimates of the 
shifts of the Eckhaus and zigzag stability boundary with decreasing system size, for several boundary conditions along the
transverse direction.  This is also checked numerically in section \ref{Lincharts}. There we also compare the analytical results on the instability limits with numerical results for the case when amplitude-suppressing boundary conditions are used in the transverse direction.

 The results in section\,,\ref{sec:Lin_stab_stripes} describe that in small systems one finds remarkable broadenings of the stable wavenumber band by reducing the system size.
   Also, stripes in two spatial dimensions behave quasi-one-dimensional  in rather narrow systems. In contrast, the evolution of a stripe pattern from an unstable to a stable wavenumber already in medium-size systems depends significantly on the nature of the boundary conditions, as we show with exemplary simulations in \secref{Numresult}.  A summary and conclusions are given in section \ref{conclu}.


\section{Models}
\label{sec: model} 
 
Generic properties  of stripe patterns just above threshold   can be described by  models such as the isotropic Swift-Hohenberg (SH) model \cite{Hohenberg:1977.1,CrossHo,Cross:2009}. 
It contains the characteristic wavenumber $q_0$  of the pattern and allows in two spatial dimensions  also the modeling of spatial variations of the local wavevector of periodic patterns.
 Essential universal properties of periodic patterns are also captured by the dynamics of the pattern envelope varying slowly on the wavelength of the pattern.\cite{Newell:1969.1,CrossHo,Cross:2009}. The dynamical equation for the envelope in isotropic systems is described by the so-called universal Newell-White-Segel equation (NWSE) 
\cite{Newell:1969.1,Segel:69.1,Newell:1993.1,CrossHo,Cross:2009}, which can be derived
 from the SH model as well \cite{CrossHo}. We use here both complementary model descriptions of stripe patterns.

\subsection{Swift-Hohenberg model}

The rotational invariant Swift-Hohenberg model for the scalar field  $u({\bf r},t)$   
above  supercritical bifurcations is\cite{Hohenberg:1977.1,CrossHo,Cross:2009},
\begin{align}
\label{SHbasic}
\partial_t u = \varepsilon u -(q_0^2+\nabla^2)^2u-u^3\,,
\end{align}
with the control parameter $\varepsilon$  and the intrinsic wavenumber  $q_0$. 
Near the threshold of  periodic patterns ($\varepsilon_c=0$) the solutions of the SH-model
can be also expressed in terms of a slowly varying amplitude $A(x,y,t)$ as follows:
\begin{align}
\label{SHmode}
 u(x,y,t)=A(x,y,t) e^{iq_0 x} + A^\ast(x,y,t) e^{-iq_0 x}  + h.o.t.\,.
\end{align}

\subsection{Newell-Whitehead-Segel equation}
\label{pertexp}

The  dynamical equation for the patterns envelope $A(x,y,t)$ can be derived from the SH model  as well as other pattern forming systems, such as from the basic equation of thermal convection
in liquids. 
Through a systematic multiscale perturbations analysis around the  onset of periodic patterns and using the property that
 the envelope $A(x,y,t)$ varies slowly on 
the scale of a wavelength of the pattern, $2\pi/q_0$,  \cite{Newell:1969.1,Segel:69.1,Newell:1993.1,CrossHo,Cross:2009}
one obtains the so-called Newell-Whitehead-Segel equation (NWSE)
for the amplitude $A(x,y,t)$:
\begin{align}
\label{NWSE}
\tau_0 \partial_t A = \varepsilon A +\xi_0^2(\partial_x -\frac{i}{2q_0}\partial_y^2)^2 A - g_0 |A|^2A\,.
\end{align}
The systems specific properties are covered by the coefficients, i.e. the values and their meanings depend on the
system but the form of the NWSE is universal.   If \eqref{NWSE} is 
derived from the SH equation \eqref{SHbasic}  one obtains
 $\tau_0=1$, $\xi_0=2q_0$ and $g_0=3$.
The universal NWSE with the  coefficients corresponding to the SH model reproduces stability ranges of stripe pattern near the threshold  as well, as exemplarily shown in Appendix \ref{AppNWSESHCOMP}.

\subsection{Boundary conditions}
\label{sec:bound}
Here, the  nonlinear stripe patterns are investigated in rectangular areas with respect to 
four different boundary conditions. 
One type are periodic boundary conditions (PBC) 
\begin{align}
\label{PBC}
 u(x=0,y)&= u(x=L_x,y) \,,   \qquad\quad (\mbox{x-PBC})~\,\\
  u(x,y=0)&=u(x,y=L_y) \,. \qquad\quad~ (\mbox{y-PBC})\,
\end{align}
PBC take just into account finite-size effects and the possible wavelengths  can take only discrete values.
We also consider Neumann boundary conditions, which are also known as 
 no-flux boundary conditions:
\begin{align}
\label{BCI}
(\vec{n}\cdot \nabla)u=0=(\vec{n}\cdot \nabla)^3 u\,. \qquad (\mbox{BCI})\,
\end{align}
Vanishing amplitude and curvature normal to the rectangular boundary is implemented by he third type of boundary conditions considered here:
\begin{align}
\label{BCII}
u=0=(\vec{n}\cdot \nabla)^2 u\,. \qquad \qquad~~ (\mbox{BCII})\,
\end{align}
The fourth type of boundary conditions for the field $u(x,y,t)$,
\begin{align}
\label{BCIII}
\left. u(x,y)\right|_{y=0,L_y}=0=\partial_y \left. u(x,y)\right|_{y=0,L_y}\, \qquad (\mbox{BCIII})
\end{align}
is only used here along the long sides at $y=0,L_y$.
These boundary conditions are suitable for  example for stripe patterns in  thermal convection in finite boxes, to model
so-called no-slip boundary conditions for the flow velocity.
Effects of the boundary conditions  BCIII in both directions in a two-dimensional systems are investigated for the SH model 
also in Ref.~\cite{Greenside:84.1}.
\subsection{Stationary single-mode solutions of the SH model}
%
Perturbations $u=\bar u \exp(\sigma t+i {\bf q}\cdot {\bf r})$ with ${\bf q}=(q,p)$ of 
 the basic state  $u=0$ of the SH equation grow beyond the so-called neutral curve
\begin{align}
\label{eq:neutralSH}
\varepsilon_0(q_n,p_m) = (q_0^2-q_n^2-p_m^2)^2 \,.
\end{align}
Note, that in finite systems only discrete values $q_n$ and $p_m$
match into the system  for the boundary conditions
PBC, BCI and BCII. For PBC 
one has $q_n=n2\pi/L_x$ and $p_m=m2\pi/L_y$. For BCI and BCII the discretization steps are half the size:
 $q_n=n\pi/L_x$ and $p_m=m\pi/L_y$.
Consider  stripe perturbations  with periodicity in the  $x$-direction and taking into account the boundary conditions 
one has: 
\begin{align}
\label{growmode}
 u&=\bar u \exp(\sigma t) \cos(q_nx)\,, \qquad \mbox{(BCI)} \nonumber \\
 u&= \bar u  \exp(\sigma t) \sin(q_nx)\,. \qquad \mbox{(BCII)} 
 \end{align}
  The growth rate vanishes along $\varepsilon_0$ and
beyond $\varepsilon_0(q)$ the perturbations grow up to a saturation amplitude.

With $\sigma=0$ and $\bar u= 2A_s$ in \eqref{growmode} and for PBC or BCI in the $y$-direction  one
obtains after projection of the SH equation (\ref{SHbasic})
 onto $\cos(q_nx+\varphi)$ with $\varphi=0$ for BCI, $\varphi=\pi/2$ for BCII 
 and arbitrary $\varphi$ for PBC in the $x$-direction
the  expression for the following amplitude of stripes:
\begin{align}
\label{eq:As}
A_s^2 =\left[\varepsilon-(q_0^2-q_n^2)^2\right]/3 \,.
\end{align}
 
 The determination of the threshold and the amplitude in the case of BCIII requires essentially a numerical approach as in Ref.\cite{Greenside:84.1} and 
the onset of the periodic pattern  takes place at higher values of $\varepsilon$. In the case of BCII in  the $y$-direction and the stripe axis parallel to $y$ the nonlinear solution has to be determined numerically as well.

 \section{Stability boundaries of stripes \label{sec:Lin_stab_stripes}}
The linear stability of stripe solutions of the SH model in rectangular domains 
can be studied analytically for periodic boundary conditions 
in \eqref{PBC}, the no-flux boundary conditions \eqref{BCI}
 and for BCII-type boundary conditions in \eqref{BCII}. 
 This is described in this section and also includes the analytical determination 
 of the Eckhaus stability boundary of stripe solutions  as well as the zigzag stability boundary for stripes.
 Solutions of the Newell-Whitehead-Segel equation and their stability are
 delineated in Appendix \ref{appNWSE} including a comparison with the following results obtained for the SH model.

 \subsection{The zigzag instability for  the SH model\label{sec:SHZIG}}
 To investigate the zigzag instability we add a small perturbation $v(x,y,t)$ to the stripe solution  $u_0=A_s \exp(iq_nx)+cc$: $u=u_0+ v$. A
linearization of \eqref{SHbasic} with respect to $v$ gives the linear equation,
\begin{align}
\label{eq:lin_shv}
 \partial_t  v= [\varepsilon - (q_0^2+\nabla^2)^2 -3u_0^2] v   +{\cal O}(v^2)\,.
\end{align}
It is solved analytically  by the ansatz
\begin{align}
\label{eq:zig_pert}
 v&=  e^{\sigma t} \cos(q_nx+\varphi) \left(e^{ip_jy} v_1 + e^{-ip_jy}v_2\right) 
\end{align}
with discrete  wavenumbers $q_n$  and the following boundary types in the  $x$-direction, PBC, BCI ($\varphi=0)$ and BCII ($\varphi=\pi/2$) 
and discrete values $p_j$ in  the $y$-direction for PBC  and BCI  boundary conditions .
The only difference for the boundary conditions are the allowed discrete
values of $q_n$ and $p_j$. These are
 $q_n= n \pi/L_x$ for BCI, BCII and $q_n=2n\pi/L_x$ for PBC in the 
 $x$-direction and analogously in the $y$-direction for BCI $p_j=j\pi/L_y$ 
 and for PBC $p_j=j2\pi/L_y$.
The result contradicts the recent claim in Ref.\cite{Yochelis:2020.1}  that the ansatz for the zigzag instability
in \eqref{eq:zig_pert} does not hold in the case of no-flux  boundary conditions in the $x$-direction. 
Moreover, in Ref. \cite{Yochelis:2020.1} it was claimed, that the perturbation $v(x,y,t)$ 
must become small near $x=0,L_x$, which is definitely not imposed by no-flux boundary conditions BCI.

Collecting  the linear independent contributions $ \propto e^{ i(q_nx \pm 
p_jy)}$ in \eqref{eq:lin_shv} gives two coupled homogeneous equations
for $v_1$ and $v_2$ with the solubility condition
\begin{align} 
 \begin{vmatrix}
  {\cal L} -6 A_s^2 \qquad & -3 A_s^2~~\\
 -3 A_s^2 ~~~&   {\cal L} -6 A_s^2
 \end{vmatrix}
 =0\,,
 \end{align}
where the linear operator ${\cal L}$ is given by 
 \begin{align} 
  {\cal L}= \varepsilon -\sigma -(q_0^2-q_n^2-p_j^2)^2\,.
 \end{align}
With the stationary amplitude $A_s(q_n)$ from \eqref{eq:As} the growth rate 
takes the following form:
\begin{align}
\label{eq:sig_qp_zig}
 \sigma= p_j^2\left[2\left(q_0^2-q_n^2\right)-p_j^2\right] \,.
\end{align}
For a nonlinear periodic solution of wavenumber $q_n$ this growth rate
$\sigma$ of the perturbation $v$ with the transversal wavenumber $p_j$
becomes positive  when the wavenumber $q_n$ becomes smaller than $q_{zz}$ at
the zigzag stability boundary:
\begin{align}
\label{SHzigzagb}
 q_n < q_{\text{zz}}=\sqrt{q_0^2-\frac{p_j^2}{2}}\,.
\end{align}
The shift of $q_{zz}$ to a value smaller than $q_0$ is determined by the smallest perturbation wavenumber 
$p_1$ that matches into the interval $[0,L_y]$.  $p_1$ is two times 
larger for periodic boundary conditions and therefore the 
shift of $q_{zz}$ away from $q_0$ is 
larger in the case of periodic boundary conditions, compared to BCI. This also means that the stable $q$ range for stripes 
is stronger enhanced for PBC than for BCI as shown in \figref{fig:Eckzig} below.

\subsection{Finite size effects on the Eckhaus instability}
The determination of the stability of periodic patterns  against longitudinal
perturbations gives the so-called  Eckhaus-boundary stability boundary  and for this
a one-dimensional analysis of \eqref{eq:lin_shv} is sufficient.
In simulations one can achieve a quasi  one-dimensional  behavior of stripe patterns 
by choosing  in simulations a small extension $L_y$ (see also \secref{Lincharts}).  
Therefore,  we choose a one-dimensional ansatz $v(x,t)$ for \eqref{eq:lin_shv}:
\begin{align}
\label{eq:longi_pert}
 v=e^{\sigma t} \left[e^{iq_n x} \left(e^{ik_j x} v_1 + e^{-ik_jx}v_2^\ast  \right)+cc  \right] \,.
\end{align}
In finite systems also the wavenumber $k_j$ of the perturbation is limited to
discrete values. For   boundary conditions BCI and BCII along $x$
one has $k_j=j\pi/L_x$ and for periodic boundary conditions  one has $k_j=j 2\pi/L_x$.
 Also the longitudinal instability is  a along wavelength instability,
i.e. growth rate becomes at first positive at the  smallest perturbation wavenumber $k_1$.
 If the perturbation $v$ is growing with a wavenumber $k_1$, then 
during the instability process either a node is added to the 
stripe solution or
removed. By using   the ansatz (\ref{eq:longi_pert}) in \eqref{eq:lin_shv} and
collecting  the  contributions $ \propto e^{ i(q_n \pm k_1) x}$ one obtains two coupled homogeneous equations
for $v_1$ and $v_2$ with the solubility condition
\begin{align} 
 \begin{vmatrix}
  {\cal L}_+ -\sigma  \qquad & -3 A_s^2~~\\
 -3 A_s^2 ~~~&   {\cal L}_- -\sigma
 \end{vmatrix}
 =0.
 \end{align}
 and the abbreviation
 \begin{align}
 \label{express_Lpm}
  {\cal L}_\pm&= -\varepsilon + \underbrace{2(q_0^2-q_n^2)^2 -(q_0^2- (q_n\pm k_1)^2)^2}_{M_\pm}\, .
 \end{align}
The growth rate of the perturbation expressed in terms of ${\cal L}_\pm$ is given by
\begin{align}
\label{sig_qp_long}
 \sigma&= \frac{1}{2} \left( {\cal L}_+ +{\cal L}_- + \sqrt{ ({\cal L}_+-{\cal L}_-)^2+36 A_s^4} \right)\,.
\end{align}
The neutral stability condition $\sigma=0$ for  periodic solution of wavenumber $q_n$ is given by 
\begin{align}
 ( {\cal L}_+ +{\cal L}_-)^2 &= ( {\cal L}_+ -{\cal L}_-)^2+36A_s^4
\end{align}
or in different form by
\begin{align}
   {\cal L}_+ {\cal L}_-&=9A_s^4\,.
\end{align}
This gives the control parameter value $\varepsilon(q_n)=\varepsilon_n$
 at the neutral stability of a stripe pattern of wavenumber $q_n$
\begin{align}
\label{Eckhaus_finite}
 \varepsilon(q_n)=\varepsilon_n=\frac{M_+M_--(q_0^2-q_n^2)^4}{M_++M_-  -2(q_0^2-q_n^2)^2}\,.
\end{align}
 For 
$\varepsilon> \varepsilon_n$ the periodic solution is stable and below 
 in the range $\varepsilon_0(q_n)<\varepsilon< \varepsilon_n$ linear unstable
with respect to  the 
  longitudinal perturbation of wavenumber $k_1$.
%

%
\section{Linear stability diagrams of stripes \label{Lincharts}}

\begin{figure}[htb]
	\includegraphics[scale=1]{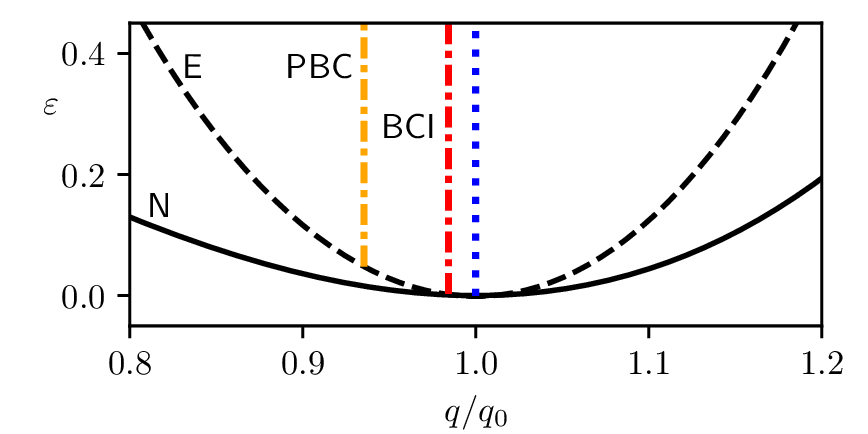}
	\caption{Shown is the  neutral curve (solid line), the Eckhaus-stability boundary (dashed line)
	and the vertical lines mark the zigzag instability $q_{zz}$ given by 
	\eqref{SHzigzagb} for different $L_y$ and for BCI and PBC in  the $y$-direction.  
	The vertical dotted line marks $q_{zz}$ for  both, BCI and PBC at $L_y=40 \lambda_0$. 
	For $L_y=2 \lambda_0$ the position of $q_{zz}$ is stronger shifted for PBC (left dash-dotted line)
	than for   BCI (right  dash-dotted line).
	\label{fig:Eckzig}
	 }
\end{figure}

Stability diagrams of stripe patterns with their wavevector ${\bf q}=(q,0)$ along the $x$ axis are presented 
in this section.  The position of the zigzag-stability boundary, $q_{zz}$, and the Eckhaus-stability boundary (E) of stripe patterns in a rectangular domain are shown in \figref{fig:Eckzig}  for two widths $L_y$, in the $y$-direction either periodic (PBC) or no-flux boundary conditions (BCI)  and no-flux boundary conditions at $x=0,L_x$.
The solid line N in \figref{fig:Eckzig} is the neutral curve described by \eqref{eq:neutralSH}. 
 The vertical dotted line in \figref{fig:Eckzig}
 marks the position of the zigzag-stability boundaries $q_{zz}$ obtained via \eqref{SHzigzagb}
  for a broad rectangle with $L_y=40\lambda_0$
 and for BCI and PBC
at $y=0,L_y$. The $q_{zz}$  for both cases is indistinguishable.  In a narrow system with $L_y=2\lambda_0$ the zigzag-stability boundary for PBC in  the $y$-direction in \figref{fig:Eckzig} is  about four times as far  shifted to the left than for  BCI. The reason, the smallest wavenumber of a perturbation of the periodic stripe solution in a system of width $L_y$ is for PBC twice as large as for BCI, i.e., $p_1^{PBC} =2 p_1^{BCI}$ and its square  $p_1^2$  contributes to  $q_{zz}$
in \eqref{SHzigzagb}. Thus, the location of the zigzag stability boundary depends essentially on the system extent $L_y$ and on the boundary conditions in the $y$-direction of the stripe axis, but not on the boundary conditions perpendicular to the stripe wavevector
in  the $x$-direction.

\begin{figure}[htb]
	\includegraphics[scale=1]{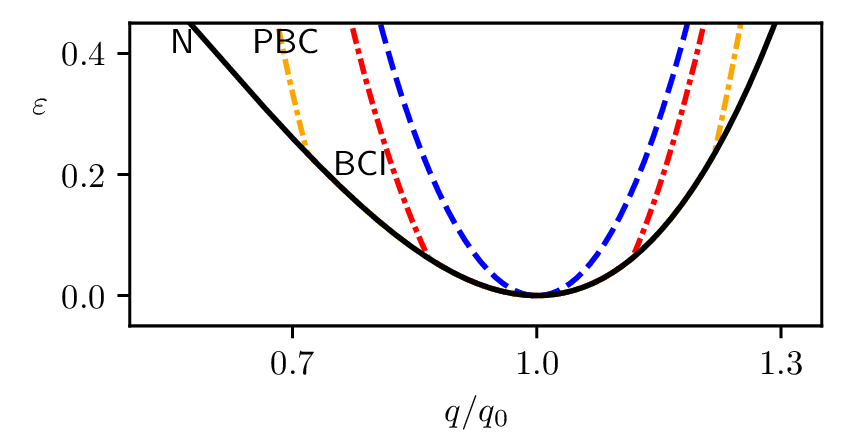}
\vspace{-4mm}
	\caption{Shown is the ESB in the one-dimensional case for both, PBC and BCI in  the $x$-direction. In a long system  with $L_x=40\lambda_0$ the ESB's  are indistinguishable for PBC and BCI (dashed line). In a short system with $L_x=2 \lambda_0$ (dash-dotted lines) the Eckhaus stability range is broader for PBC than for BCI. The solid line N is the neutral curve.  
	} 
\label{fig:Eckhaus_finite}
\end{figure}

In  one-dimensional systems with  large $L_x$
 the Eckhaus-stability boundaries (ESB) for BCI and PBC in  the $x$-direction are
 indistinguishable  and
 given  by the dashed line  in  \figref{fig:Eckhaus_finite}. 
 The location of the ESB depends  on the length  $L_x$ and also on boundary conditions in the $x$-direction. To illustrate this, we show in \figref{fig:Eckhaus_finite} also the Eckhaus boundary for a short system with only two periodic pattern units in the system. For this purpose, we consider the case of a pattern of wavenumber $q$ in a system of suitable length $L_x= 2 (2\pi/q)$ for the two boundary conditions PBC and BCI
at $x=0,L_x$.
The ESB is determined by \eqref{Eckhaus_finite}, but with $k_1= \pi/L_x$ for BCI and
$k_1=2\pi/L_x$ for PBC. 
For both short systems, the ESB intersects the neutral curve, as shown in \figref{fig:Eckhaus_finite}.
For $\varepsilon$ below this intersection the periodic patterns are  stable for all $q$ between  neutral curve, i.e. for all stripe patterns that emerge.
The dependence of the Eckhaus boundary on the system length
was already recognized in Ref.~\cite{Zimmermann:85.1} and this length dependence was also confirmed in experiments on Taylor vortex flow in Ref.~\cite{Dominguez-Lerma:86.2}.  
 The ESB is similar as for the zigzag boundary 
 for PBC shifted further  from the Eckhaus boundary for very long systems  than for the boundary condition BCI.
The difference is again caused by the different value of the perturbations wavenumber, similar as for the zigzag instability.

 The wavenumber along the Eckhaus curve, as for example along  the dashed curve in 
 \figref{fig:Eckhaus_finite}, we call $q_E(\varepsilon)$. Its deviation from the preferred wavenumber $q_0$ is $Q_E=q_E-q_0$. Defining analogous $q_N(\varepsilon)$ as the wavenumber along the neutral curve and its deviation from the preferred wavenumber $Q_N=q_N-q_0$, the ratio between both deviations follow near the threshold ($\varepsilon \ll 1$) the universal law of stripe patterns in one spatial dimension \cite{Newell:1969.1,Zimmermann:85.1,Cross:2009}:
\begin{align}
\frac{Q_E}{Q_N}=\frac{1}{\sqrt{3}}\,.
\label{eq:eckhaus_infinite}
\end{align}
The ratio becomes larger in systems of finite extension $L_x$. In addition, the ratio
 depends on $\varepsilon$, $L_x$
 and on the boundary conditions in the longitudinal $x$-direction.
\begin{figure}[htb]
	\includegraphics[scale=1]{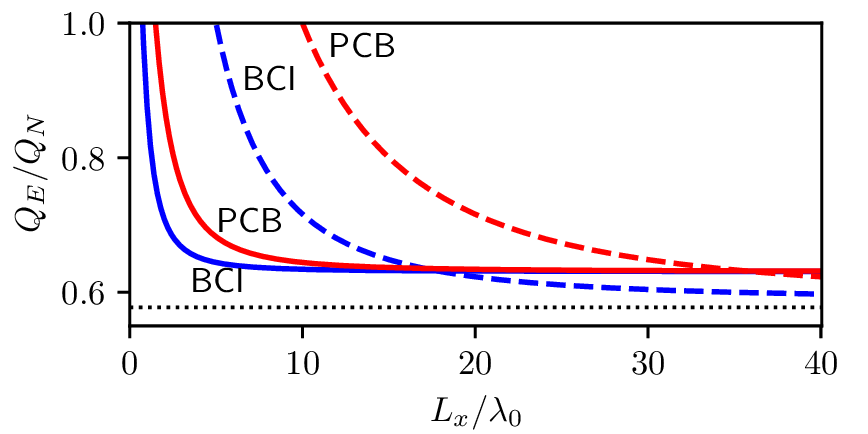}
	\caption{Shown is  the ratio $Q_E/Q_N$ between the half-width of the Eckhaus-stability band, $Q_E$,
	and the half-width of the neutral curve, $Q_N$,  as function of the system length $L_x/\lambda_0$ for the two boundary conditions  PBC and  BCI and
		 for the control parameter $\varepsilon=0.01$ (dashed lines) and  $\varepsilon=0.4$ (solid curves).
		 \label{fig:QEQN}
		 }
\end{figure}
We determine via \eqref{Eckhaus_finite} the wavevector along the Eckhaus boundary,  $q_E(\varepsilon,k_1)$, as function
of $\varepsilon$ 
 with $k_1=\pi/L_x$ for BCI (resp. $k_1=2 \pi/L_x$ for PBC). The ratio between $Q_E(\varepsilon,k_1)$
 and  $Q_N=q_N-q_0$  is shown for the SH model in \figref{fig:QEQN} as function of
 $L_x$ for two different values of $\varepsilon$ and two boundary conditions.  
For an infinite system near threshold this ratio is given by \eqref{eq:eckhaus_infinite}, which is indicated by the dotted horizontal line in \figref{fig:QEQN}. 
 For PBC the ESB is shifted further  away  from the value for infinite systems in \eqref{eq:eckhaus_infinite}  and closer to the neutral curve 
$\varepsilon_0(q)$ than for BCI. This means the ESB reaches the neutral curve (see also \figref{fig:Eckhaus_finite}) and thus the largest possible value $Q_E/Q_N=1$ for a chosen $\varepsilon$ already at larger system lengths $L_x$ than for no-flux boundary conditions BCI.

\begin{figure}[htb]
	\includegraphics[scale=1]{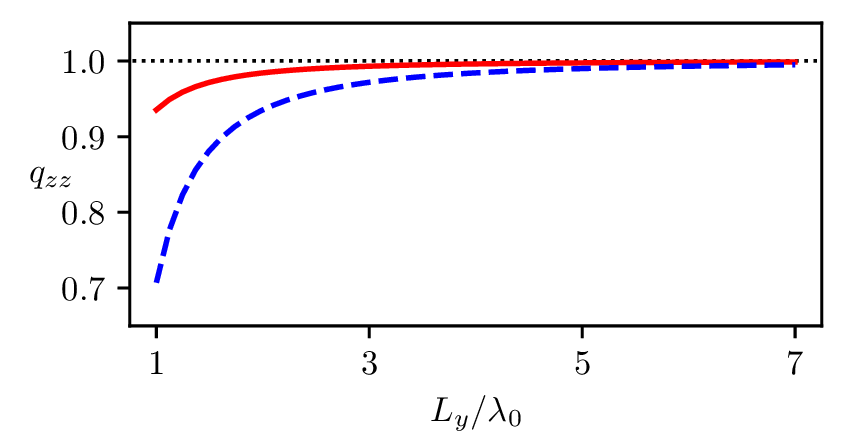}
	\caption{The zigzag stability boarder $q_{zz}$, given by \eqref{SHzigzagb}, is shown as a function of the system width $L_y$ for no-flux boundary conditions (BCI) (solid line) and for periodic boundary conditions (dashed line) in the $y$-direction.
	\label{fig:qly}
	}
\end{figure}

The  value $q_{zz}$ at the  $\varepsilon$-independent zigzag-stability boundary 
has different values for different boundary conditions in  the $y$-direction as indicated in 
  \figref{fig:Eckzig}. The dependence of $q_{zz}$
  on the system width $L_y$ is shown in \figref{fig:qly}
  for the two boundary conditions PBC and BCI in  the $y$-direction.
  $q_{zz}$ decreases with decreasing $L_y$ in both cases, but
   stronger for PBC. The reason is again the larger perturbation wavenumber $k_1$ in \eqref{express_Lpm} for PBC.

\begin{figure}[htb]
	\includegraphics[scale=1]{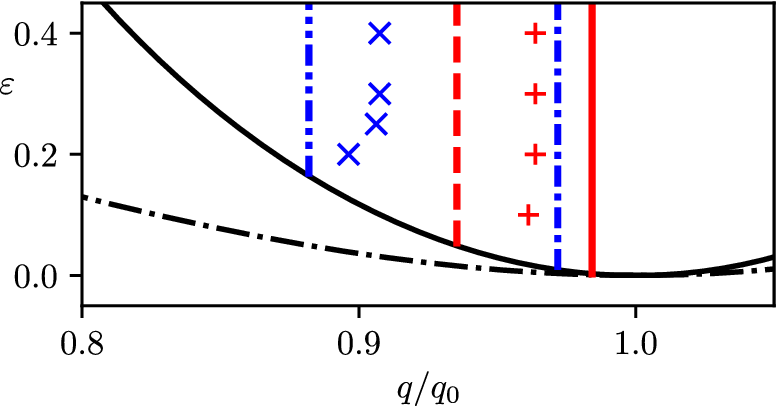}
	\caption{In a  long system with $L_x=40\lambda_0$, 
	   the ESB (solid line)
	and  the zigzag stability lines are shown for different boundary conditions 
in  the $y$-direction and for two  widths $L_y=2\lambda_0,1.5 \lambda_0$.
Width $L_y=2\lambda_0$   (red): The vertical solid line marks the zigzag stability boundary  $q_{zz}$ for BCI, the vertical dashed line $q_{zz}$ for PBC and the $+$-symbols mark $q_{zz}$ for BCII. 
Width $L_x=1.5\lambda_0$ (blue): The vertical dash-dotted line
gives $q_{zz}$  for BCI, dash-dot-dotted line for PBC and the crosses $\times$ mark $q_{zz}$ for  BCII.
\label{fig:zigzag_all}	
	}
\end{figure}

  A consequence of the $L_y$ dependence of $q_{zz}$ in \figref{fig:qly} is further illustrated in 
 \figref{fig:zigzag_all}. Shown is the neutral cuve (dashed-dotted), the Eckhaus boundary (solid line)
  and the zigzag instability for the two lateral extensions $L_y=2\lambda_0,~1.5 \lambda_0$ 
  and three different boundary conditions in  the $y$-direction.  For the wider system with $L_y= 2 \lambda_0$
  the zigzag instability 
  $q_{zz}$ for  BCII ($+$ symbols)  is located between $q_{zz}$
  for BCI  (vertical solid line)  and $q_{zz}$ for PBC (vertical dashed line).
  Also for the narrower system with  $L_y= 1.5 \lambda_0$ the zigzag instability 
  $q_{zz}$ for the type BCII boundary condition ($\times$ symbols)  is between $q_{zz}$
  for BCI boundary conditions (vertical dashed-dotted line)  and $q_{zz}$ for PBC (dashed-dot-dotted line).
While the results for BCI and PBC   are determined by the expression in \eqref{SHzigzagb}, the  
zigzag instability  for y-BCII are determined via simulations.
  In this sense the analytical formula of $q_{zz}$ for BCI and PBC gives a reasonable 
  estimate about the location of the zigzag boundaries for further boundary conditions in  the $y$-direction.  

In \figref{fig:zigzag_all} the zigzag boundary crosses 
for different boundary conditions  the Eckhaus-boundary (solid curve) at different values of $\varepsilon$.
For control parameter values below this intersections of the zigzag and the Eckhaus boundary the systems behaves quasi-one dimensional. I.e. below these $\varepsilon$ the zigzag (ZZ) instability becomes irrelevant,  because the
Eckhaus instability sets in earlier.

\begin{figure}[htb]
	\includegraphics[scale=1]{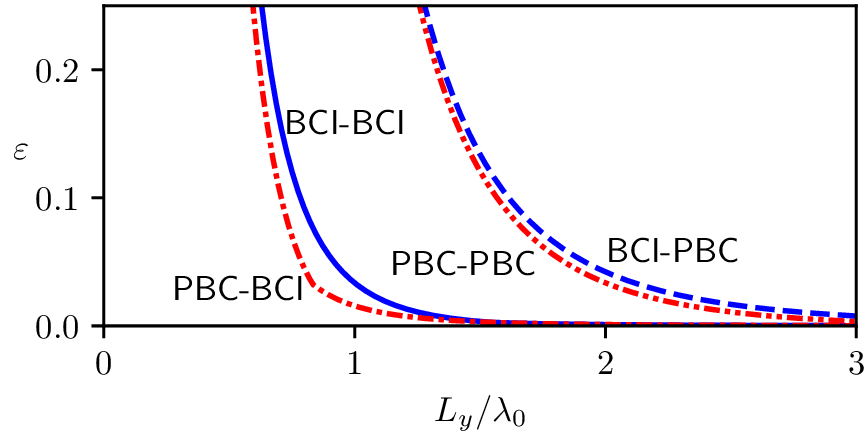}
	\caption{For a short system length $L_x=3 \lambda_0$ and  different combinations of BCI and PBC in the $x$- and  the $y$-direction, the $\varepsilon$ position of the intersections between   the ESB and the zigzag boundary is shown as function of the system width $L_y$.  The first part of the curve designation describes the boundary  in  the $x$-direction.
	}
\label{fig:epsly}
\end{figure}

Since the ESB depends for small system lengths $L_x$ on the boundary condition in  the $x$-direction,
the crossing of the ZZ and the Eckhaus stability boundary depends on the boundary conditions in the $x$ and $y$ direction. Therefore, the transition to a quasi one-dimensional behavior of stripe patterns depends on the control parameter $\varepsilon$, the system size and the used boundary condition in each direction. 
This is shown in \figref{fig:epsly} for $L_x=3 \lambda_0$,  where the first part of the curve label refers to the boundary condition in  the $x$-direction and the second part refers to the boundary condition in  the $y$-direction.
Below these four curves
in \figref{fig:epsly} the zigzag instability becomes irrelevant for these systems sizes and boundary conditions and the stripes behave one dimensional.
Conversely, 
the zigzag instability limits above these curves in \figref{fig:epsly} the stability stripe pattern in the range $q<q_0$
for all the considered systems sizes and boundary conditions  considered in this work, which is in contrast to Reference \cite{Yochelis:2020.1}.

\section{Examples of the nonlinear evolution of unstable stripes in finite systems}
\label{Numresult}

In this section, we exemplify how stripe patterns evolve  after applying small perturbations from an unstable wavenumber $q<q_0$  through nothing but a zigzag instability to stripe patterns at a stable wavenumber. We use simulations of the SH model (\ref{SHbasic}) in a rectangular domain with $L_x =40 \lambda_0$ and $L_y = 20 \lambda_0$ and choose several  different boundary conditions along the sides of the rectangle.

\begin{figure}[htb]
	\includegraphics[scale=1]{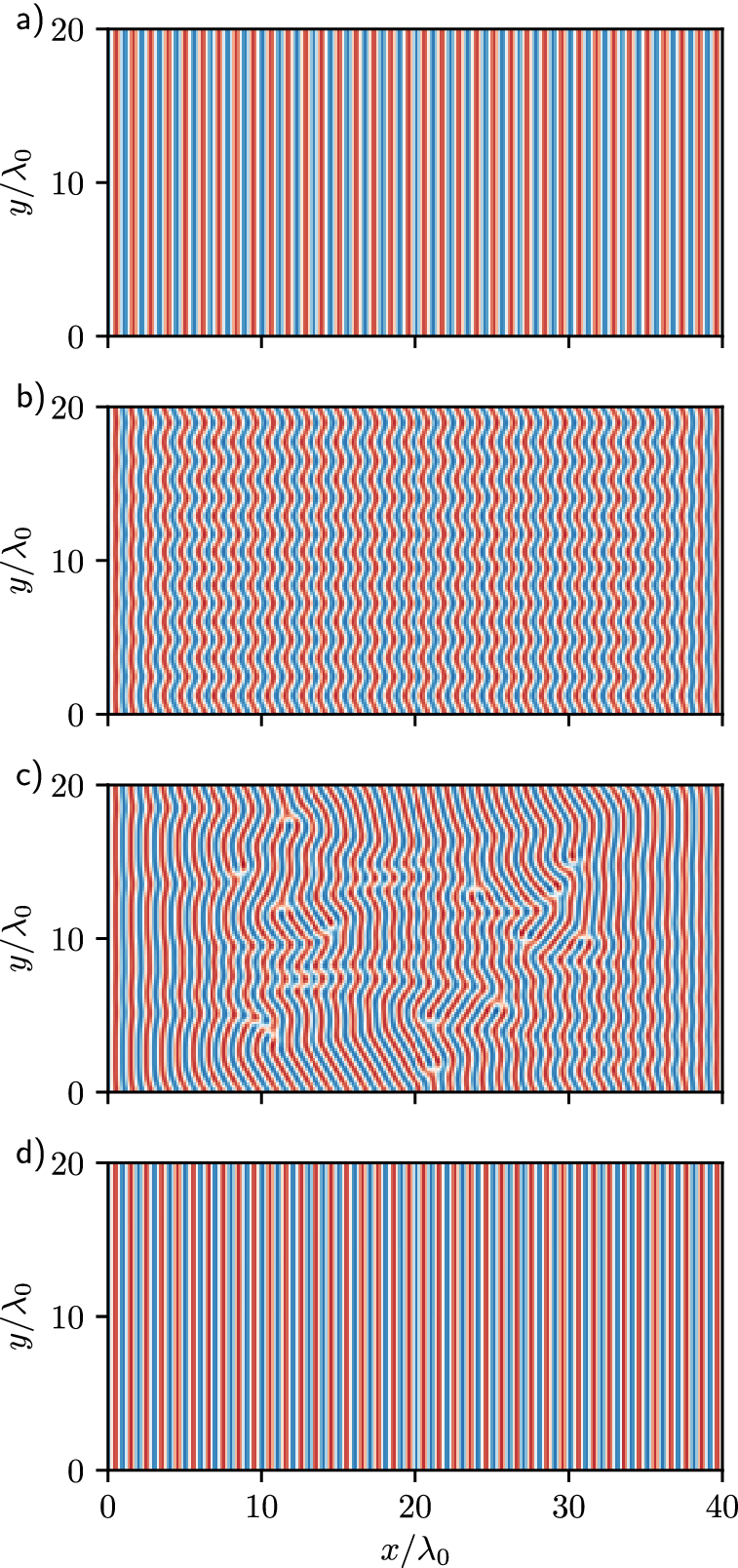}
	\caption{Shown are four snapshots of a simulation of the SH model in a rectangular domain of the side lengths 
	$L_x=40 \lambda_0$ and  $L_y=20 \lambda_0$; color scale represents the minimal (blue) and the maximal (red) values of $u$;  no-flux boundary conditions (BCI)
	along the four boundaries and at $\varepsilon=0.4$. In  (a) the initial 
	solution with the wavenumber $q_s=0.9$ at time $t=0$ is shown,  
	which is undulated by the zigzag instability in (b) at $t= 10^3$. 
	The resulting defects in (c) at $t=7.5\cdot 10^3$ heal rapidly due to the fixed phases at the edges and result in the solution with wavenumber $q=1.0$ at $t=66\cdot 10^3$ in (d).}
\label{app_fig_noflux}
\end{figure}
\begin{figure}[htb]
	\includegraphics[scale=1]{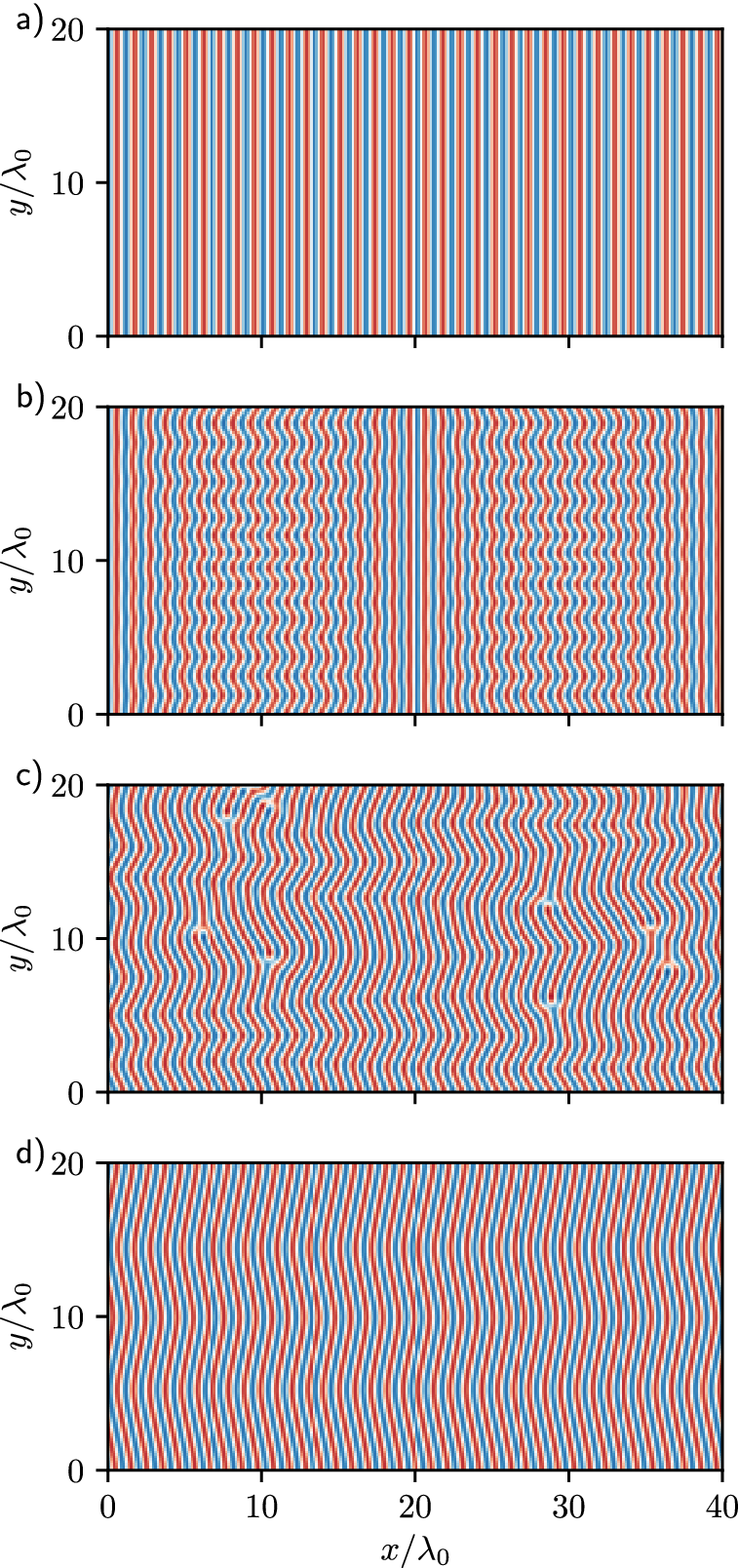}
	\caption{Shown are four snapshots of a simulation of the SH model in a rectangle and the  same parameters as
	 in \figref{app_fig_noflux}. Here,   the no-flux boundary conditions are replaced
	by  periodic boundary conditions (PBC) in  the $x$-direction: starting from a solution with  $q_s=0.9$ in (a) at 
	time $t=0$, the solution becomes modulated at  $t= 2\cdot 10^3$ in (b). 
	The defects appearing at $t=3.5\cdot 10^3$ in (c) are eliminated during further evolution and result in a solution with $q=1.0$ at $t=171\cdot10^3$.
	}
\label{app_fig_peri}
\end{figure}
In \figref{app_fig_noflux} and \figref{app_fig_peri} we started simulations  in  the rectangular domain by using  a  periodic initial pattern with $\varepsilon=0.4$ and a 
wavenumber $q=0.9$ which produce 36 stripes in the system. In \figref{app_fig_noflux} no-flux boundary conditions (BCI) are used 
along all four sides, whereas in \figref{app_fig_peri} we  replaced BCI by periodic boundary conditions
in  the $x$-direction. For the chosen system extensions  the 
zigzag-stability boundaries is in both cases  nearly indistinguishable at $ q_{zz} \lesssim q_0$
as shown in \figref{fig:Eckzig}, i.e.  
the starting wavenumber $q=0.9$ is far in the unstable range. While the two different boundary conditions
leave the zigzag-stability boundary nearly untouched for larger $L_y=20\lambda_0$, 
the temporal evolution of the stripe patterns from an unstable to a stable wavenumber $q$
differs. Away from the boundaries at $x=0,L_x$
the unstable stripes with $q_0=0.9$ become according to the zigzag instability undulated, as can be seen in 
 \figref{app_fig_noflux}b)  and \figref{app_fig_peri}b). 
 For periodic boundary conditions at $x=0,L_x$ 
  these stripe undulations occur also at the boundaries as shown in  \figref{app_fig_peri}c).
  No-flux boundary conditions  fix the phase of the stripe pattern with its maximum or minimum $u(x,y,t)$ 
   at $x=0,L_x$. Therefore 
  no undulations evolve at and near the boundaries, as indicated  in  \figref{app_fig_noflux}c). According
  to the BCI induced constraint on the phase    near the boundaries one has for BCI at $x=0,L_x$ 
  a shorter transient time to reach finally a  
   stable straight stripe pattern 
  with $q_0 \lesssim q$  as in indicated
  by \figref{app_fig_noflux}d) and \figref{app_fig_peri}d). In both cases one ends up with a state composed 
  of  $40$ pattern units, i.e.  with $q_0 \lesssim q$. 
   When we start simulations for both boundary conditions with
an initial solution of wavenumber  $q=0.975$, which corresponds to $39$ pattern units in the system,
 the pattern evolves again
to a state of $40$ units with $q_0\lesssim q$.

  \begin{figure}[htb]
	\includegraphics[scale=1]{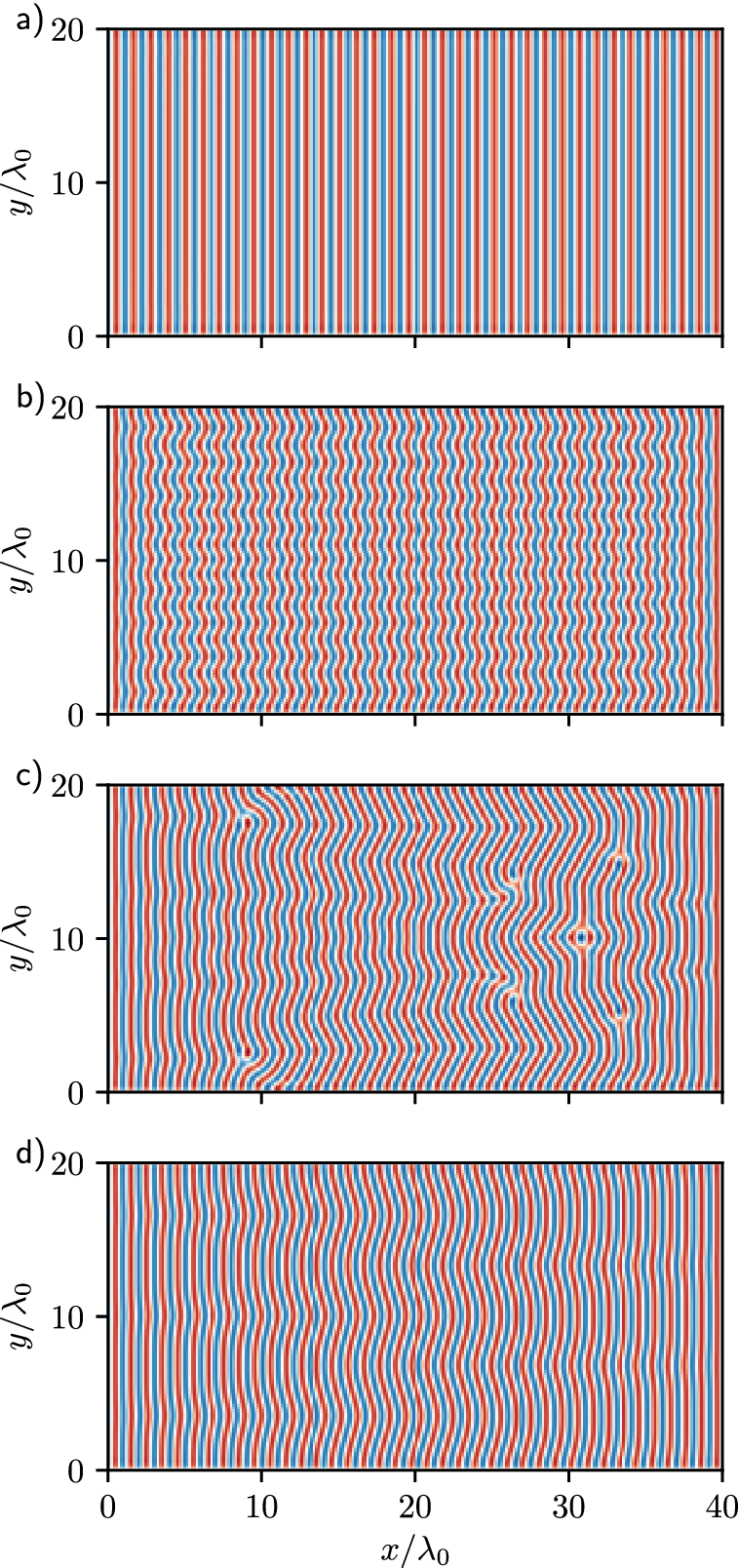}
	\caption{Shown are four snapshots of a simulation in a rectangle for the same parameters as in 
	\figref{app_fig_noflux}, but with  type II boundary conditions (BCII) in  the $y$-direction: 
	a) Initial solution with 
	wavenumber $q_s=0.9$ at $t=10$, b) $t= 500$,
	c) $t=5.4\cdot 10^3$ and d) $t=109\cdot10^3$ a stable  solution with 39 periods  and $q=0.975$.}
\label{sim_BCII}
\end{figure}
  \begin{figure}[htb]
	\includegraphics[scale=1]{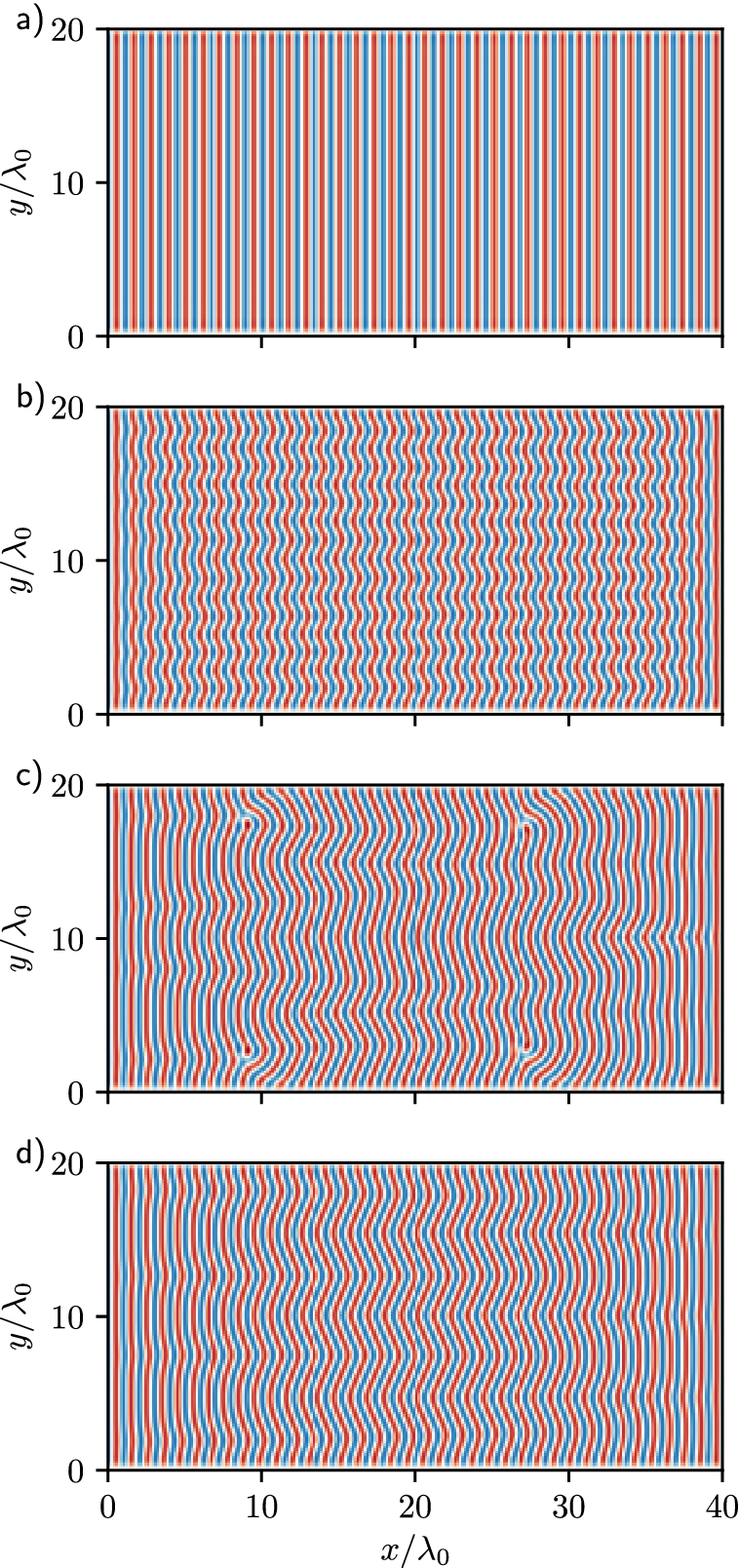}
	\caption{Shown are four snapshots of a simulation in a rectangle for  the same parameters as in 
	 \figref{app_fig_noflux}, but  type III boundary conditions (BCIII) in  the $y$-direction: a) Initial solution with 
	wavenumber $q_s=0.9$ at $t=10$, b) $t= 500$ and
	c) $t=5\cdot 10^3$ and in d) a  solution with 38 stripes and $q=0.95$ so far.
	}
\label{sim_BCIII}
\end{figure}

 In \figref{sim_BCII} and \figref{sim_BCIII} we started simulations   in  a rectangular area 
 with the same initial wavenumber $q=0.9$ and $\varepsilon$ as in \figref{app_fig_noflux}. However, we replaced 
 at $y=0,L_y$ no-flux boundary conditions (BCI) by the boundary conditions of type
 BCII in \figref{sim_BCII} and by type BCIII boundary condition in \figref{sim_BCIII}.
 As indicated in \figref{fig:zigzag_all}, the position of the zigzag-stability boundary $q_{zz}$  is in the case of 
 BCII boundary conditions at $y=0,L_y$ stronger influenced in rather narrow systems than for no-flux boundary conditions, which  is also true for BCIII boundary conditions.
 
 However, the starting wavenumber $q_s=0.9$ is again
 considerably below $q_{zz}$ of PBC, which causes the 
 strongest shift.  
One can recognize in \figref{sim_BCII} and \figref{sim_BCIII} that
the boundary conditions BCII and BCIII  suppress stripe pattern close to $y=0,L_y$. 
 Also for these boundary conditions at $y_0=0,L_y$ the unstable stripe pattern at $q_s=0.9$ becomes 
 undulated in the bulk via the zigzag instability  as indicated  in \figref{sim_BCII}b) and \figref{sim_BCIII}b). The further evolution is slightly different from the evolution shown in 
  \figref{app_fig_noflux}.  The major difference is that at a similar simulations time 
  the state in \figref{sim_BCII}d) is composed of $39$ periodic units and in \figref{sim_BCIII}d)
  by $38$ periodic units.

\section{Summary  and conclusions \label{conclu}}

 We investigated  finite  sizes effects on the multistability of supercritical bifurcating stripe patterns in rectangular domains using the generic Swift-Hohenberg model and the universal Newell-Whitehead-Segel equation. In two-dimensional extended isotropic systems, the wavenumber range of stable periodic patterns is limited by the longitudinal Eckhaus instability and the transverse zigzag instability.  
We show analytically and numerically for different combinations of boundary conditions along 
the edges of a rectangular domain that the range of wavenumbers  for stable stripes increases  with a reduction of the system size. 

Note, also in finite systems the Eckhaus and zigzag instabilities remain the instabilities limiting the stable
 wavenumber range of stripe pattern for different boundary conditions. 
 The zigzag instability remains the primary transversal instability also for no-flux boundary conditions 
 in the longitudinal direction of stripe pattern. It is not 
 replaced by another primary instability as recently claimed in Ref.~\cite{Yochelis:2020.1}.

 The enlargement of the stable wavenumber range of stripe patterns by the system size reduction is based
 on the following insights. The Eckhaus and zigzag instabilities are long-wavelength instabilities. By decreasing the system size, their destabilizing long-wavelength modes are increasingly suppressed. 
 For periodic boundary conditions, an entire wavelength of a destabilizing mode must fit into the system, while for no-flux boundary conditions, for example, only half a wavelength of the destabilizing mode must fit into the finite system. That is, the smallest wavenumber $k$ of the perturbation is twice as large for periodic boundary conditions as for no-flux boundary conditions. According to our analytical results, the zigzag stability boundary is shifted proportionally to $k^2$. 
 This means, for periodic boundary conditions in transverse direction, a reduction of the rectangle width shifts the stability boundary up to a factor of four more and increases the stable wavenumber range than for other boundary condition in transverse direction.  
The enlargement trend of the stable wavenumber range is similar for a reduction of the system length in longitudinal direction by shifting the Eckhaus boundary.

If boundary conditions that suppress the amplitude of the stripe pattern are taken in the transverse direction, the numerical results for the zigzag instability boundary lie between the analytical results for periodic and no-flux boundary conditions. This underlines the value of the presented analytical results also as an estimate for the location of the zigzag instability for other boundary conditions.

 By reducing the system width sufficiently, the zigzag instability limit shifts to smaller values of the wave number than the lower Eckhaus stability limit for not to short systems. In this case, the zigzag instability is suppressed 
 by the longitudinal instability that occurred previously. 
Below such system widths, the stripe pattern behaves quasi-one-dimensionally in two-dimensional systems. Again, the transition to quasi-one-dimensional behavior for periodic boundary conditions in the transverse direction occurs at already larger widths than for no-flux boundary conditions.
 Also for the transition to quasi-one-dimensional behavior, the analytical results for periodic and no-flux boundary conditions give a good estimate for the transition to quasi-one-dimensional behavior for other boundary conditions in transversal direction.

 In the spatiotemporal evolution from a periodic stripe pattern with a wavenumber below the zigzag instability limit to a stripe pattern with a stable wavenumber, the influence of the boundary conditions is already noticeable for medium-sized systems with about 20 periodic stripes. This is shown for various combinations of boundary conditions along the rectangular domain in Section \ref{Numresult}.  As these simulations show, in all cases the zigzag instability is the destabilizing mechanism, in contrast to the description in Ref. \cite{Yochelis:2020.1}.

The results of this work give also an estimate below which
system lengths and widths, for given values of the control parameter, any emerging periodic pattern is also stable. These
insights are important in investigations of e.g. Turing
patterns in very small systems such as cells \cite{Sourjik:2017.1}.
The here derived generic limitation of the stable wavenumber bands addresses also the so called robustness problem \cite{Maini:2012.1,Sourjik:2017.1}.

\section*{Acknowledgments}
Support by the Elite Study Program Biological Physics is gratefully acknowledged.
 
 \section*{Data Availability}
 The data that support the findings of this study are available
from the corresponding author upon reasonable request.

\appendix

\section{Newell-White-Segel equation (NWSE) and stability of stripes}
\label{AppNWSE}
For completeness we include also the stability boundaries of stripes 
determined via the NWSE in Eq.\,(\ref{NWSE}) for no-flux boundary conditoins 
and periodic boundary conditions, which we can compare with the results of the SH model.

\subsection{Linear stability of stripes within the NWSE}

The NWSE has the stationary solutions
\begin{align}
A=F_ne^{iQ_nx}
\label{eq:NWSEsol}
\end{align}
with the wavenumber $Q_n=q_n-q_0$ and the amplitude $F_n^2=(\varepsilon-\xi_0^2Q_n^2)/g_0$. 
One has $F_n^2>0$ for $\varepsilon> \varepsilon_0$
and  $F_n=0$ along the neutral curve:
\begin{align}
 \varepsilon_0={\xi_0^2}Q_N^2.
	\label{eq:neutrNWSE}
\end{align}
\label{appNWSE}
\subsection{Zigzag instability within the NWSE}
At first we investigate the zigzag instability of the stripe solution.
For this we use the ansatz $A=(F+v(y))e^{iQ_nx}$ with a small  $y$-dependent perturbation $ v(y,t)$. By neglecting higher order terms of $v$ in \eqref{NWSE}, a linear equation for $v$ results: 
\begin{align}
\label{eq:lin_NWSE_v}
 \tau_0\partial_tv=\varepsilon v+ \xi_0^2\left[iQ_n-\frac{i}{2q_0}\partial_y^2\right]^2v-g_0F^2v-F^2v^*.
\end{align}
This linear equation may be solved by 
\begin{align}
\label{eq:zig_pert_NWSE}
 v=e^{\sigma t} \left[e^{iL_jy} v_1 + e^{-iL_jy}v_2^\ast \right] \,,
\end{align}
where the wavenumber for no-flux boundary conditions at $y=0,L_y$ is $L_j=j\pi/L_y$ and for PBC $L_j=j2\pi/L_y$.
Collecting  the  contributions $ \propto e^{ \pm i L_jy}$ gives two coupled homogeneous equations
for $v_1$ and $v_2$ with the solubility condition
\begin{align} 
 \begin{vmatrix}
  {\cal L} -\sigma\tau_0 \qquad & -g_0F^2~~\\
 -g_0F^2 ~~~&   {\cal L} -\sigma\tau_0
 \end{vmatrix}
 =0
 \end{align}
 and the abbreviation
 \begin{align}
  {\cal L}= -\varepsilon -2\xi_0^2Q_n^2-\xi_0^2\left(Q_n+\frac{L_j^2}{2q_0}\right)^2 \,.
 \end{align}
Herein the growth rate of the perturbation is 
\begin{align}
 \sigma \tau_0= -\frac{L_j^2}{2q_0}\left(2Q_n+\frac{L_j^2}{2q_0}\right) 
\,.
\end{align}
The stripe solution with the amplitude  given by \eqref{eq:NWSEsol}
 is stable with respect to the perturbation in \eqref{eq:zig_pert_NWSE}
 in the range 
\begin{align}
 Q_n > -\frac{L_j^2}{4q_0}\,=Q_\text{zz}.
\end{align}

\subsection{Eckhaus-instability of stripes}

Next we investigate the stability of stripe solutions with respect to 
longitudinal perturbations in long  a quasi one-dimensional systems,
i.e. with a small width $L_y=\pi/2q_0$.

In this case we investigate with \eqref{eq:lin_NWSE_v} the dynamics of perturbations $v(x,t)$ with
respect to the solution given by \eqref{eq:NWSEsol}.
We solve the linear equation (\ref{eq:lin_NWSE_v}) with the following ansatz
\begin{align}
\label{eq:longi_pert_NWSE}
 v=e^{\sigma t}\left(e^{iK_l x} v_1 + e^{-iK_lx}v_2^\ast  \right)  \,,
\end{align}
and with the wavenumber $K_l=l\pi/L_x$. If the perturbation $v$ is growing with a wave number $K_1$, then 
during the instability process either  nodes are added to the 
stripe solution, cf. \eqref{eq:NWSEsol}, or
removed. With the ansatz (\ref{eq:longi_pert_NWSE}) in \eqref{eq:lin_NWSE_v} and
collecting  the  contributions $ \propto e^{ \pm iK_j x}$, gives two coupled homogeneous equations
for $v_1$ and $v_2$ with the solubility condition
\begin{align} 
 \begin{vmatrix}
  {\cal L}_+ -\sigma\tau_0  \qquad & -g_0F_n^2~~\\
 -g_0F_n^2 ~~~&   {\cal L}_- -\sigma\tau_0
 \end{vmatrix}
 =0
 \end{align}
 and
 \begin{align}
  {\cal L}_\pm&= -\varepsilon + \underbrace{2\xi_0^2Q_n^2 -\xi_0^2(Q_n^2\pm K_l^2)}_{M_\pm}\,. 
 \end{align}
The growth rate of the perturbation expressed in terms of ${\cal L}_\pm$ is given by 
\begin{align}
\label{sig_qp_long}
 \sigma&= \frac{1}{2} \left( {\cal L}_+ +{\cal L}_- + \sqrt{ ({\cal L}_+-{\cal L}_-)^2+4g_0^2 F_n^4} \right)\,.
\end{align}
The neutral stability $\sigma=0$ condition for the $n$-node solution supplies
\begin{align}
   {\cal L}_+ {\cal L}_-&=g_0^2F_n^4\,.
\end{align}
This gives the stability boundary $\varepsilon_n$ of the $n$-node solutions in the $\varepsilon_n-q_n$ plane 
\begin{align}
\label{Eckhaus_finite_Amp}
 \varepsilon_n=\frac{M_+M_--\xi_0^4Q_n^4}{M_++M_-  -2\xi_0^2Q_n^2}\,.
\end{align}
The $n$-node solution is stable (unstable) above (below) this curve.

 \subsection{Comparison of the NWSE to the SH model}
 \label{AppNWSESHCOMP}
\begin{figure}[htb]
	\includegraphics[scale=1]{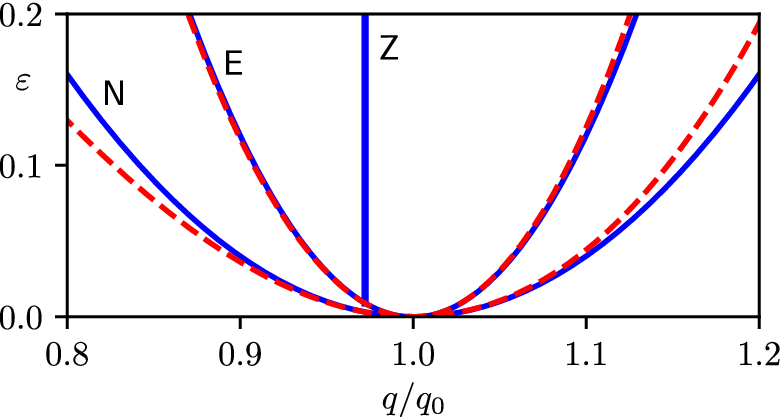}
\vspace{-4mm}
	\caption{\label{fig:ASHcompare}The neutral stability curves N, the Eckhaus-stability boundary   E and 
	the zigzag-stability boundary Z of stripes are shown as obtained 
for the  Newell-White-Segel equation (solid lines)   and the Swift-Hohenberg model (dashed lines). The zigzag-stability boundaries are indistinguishable (vertical solid line).
System size $L_x=40\lambda_0$, $ L_y=1.5\lambda_0$  and BCI in both directions.
	  }
\end{figure}

As mentioned above the NWSE can derived from the SH model for the coefficients $\tau_0=1$, $\xi_0=2q_0$ and $g_0=3$ by a weakly nonlinear analysis \cite{CrossHo}. This approximation holds near the threshold, where 
 the neutral curve of the SH model \eqref{eq:neutralSH} becomes also parabolic similar as in  \eqref{eq:neutrNWSE}. For higher values of the control parameter $\varepsilon$, this two curves N for the SH and the NWSE differ as can be seen in \figref{fig:ASHcompare}. In contrast to this  difference between the neutral curves of  the NWSE (dashed lines) and the SH model (solid lines) the Eckhaus stability boundaries (E) is nearly identical. The zigzag instability (Z) even is indistinguishable for the two models.
Therefore the qualitative and quantitative results of this section, are with universal character since the universality of the amplitude equation.

%


\end{document}